\documentclass[twocolumn,prl,showpacs,aps,superscriptaddress]{revtex4-2}

\usepackage{multirow}
\usepackage{graphicx} 
\usepackage{color}
\usepackage[latin1]{inputenc}
\usepackage{enumitem}
\usepackage[french]{babel}
\begin{document}

\date{\today}

\title{Superconductivity in cuprates governed by topological constraints}

\author{Yves Noat}

\affiliation{Institut des Nanosciences de Paris, CNRS, UMR 7588 \\
Sorbonne Universit\'{e}, Facult\'{e} des Sciences et Ing\'{e}nierie, 4 place
Jussieu, 75005 Paris, France}

\author{Alain Mauger}

\affiliation{Institut de Min\'{e}ralogie, de Physique des Mat\'{e}riaux et
de Cosmochimie, CNRS, UMR 7590,Sorbonne Universit\'{e}, Facult\'{e} des Sciences et Ing\'{e}nierie, 4 place
Jussieu, 75005 Paris, France}

\author{William Sacks}

\affiliation{Institut de Min\'{e}ralogie, de Physique des Mat\'{e}riaux et
de Cosmochimie, CNRS, UMR 7590,Sorbonne Universit\'{e}, Facult\'{e} des
Sciences et Ing\'{e}nierie, 4 place Jussieu, 75005 Paris, France}


\begin{abstract}

The remarkable universality of the cuprate $T_c$ dome suggests a very fundamental unifying principle. Moreover, the superconducting
gap is known to persist above $T_c$ in the pseudogap phase of all cuprates. So, contrary to BCS, the gap cannot be the order parameter of the transition.

In this work, we show that both the $T_c$-dome and the pseudogap line $T^*(p)$ arise from a unique and identifiable principle: the
interaction of localized `pairons' on an antiferromagnetic square lattice. The topological constraints on such preformed pairons give rise to both the $T_c$ dome and the pairing energy {\it simultaneously}. It also provides a natural explanation for the
critical doping points of the phase diagram.

The model matches perfectly both the $T^*$ and $T_c$ experimental lines, with only one adjustable parameter.
\end{abstract}

\maketitle

\subsection{Introduction}

Despite more than thirty years of intense research and many advances, a general understanding of the physics of cuprates is
still lacking. Indeed, most of the key questions remain to be answered or clarified: \vskip 2mm
\begin{enumerate}[label=\roman*)]
\item What is the pairing mechanism?
\item What is the nature of the SC condensation?
\item What is the nature of the pseudogap phase and its connection with the SC state?
\item What is the physical origin of the critical doping points?
\end{enumerate}

\begin{figure}
\includegraphics[width=6.2 cm]{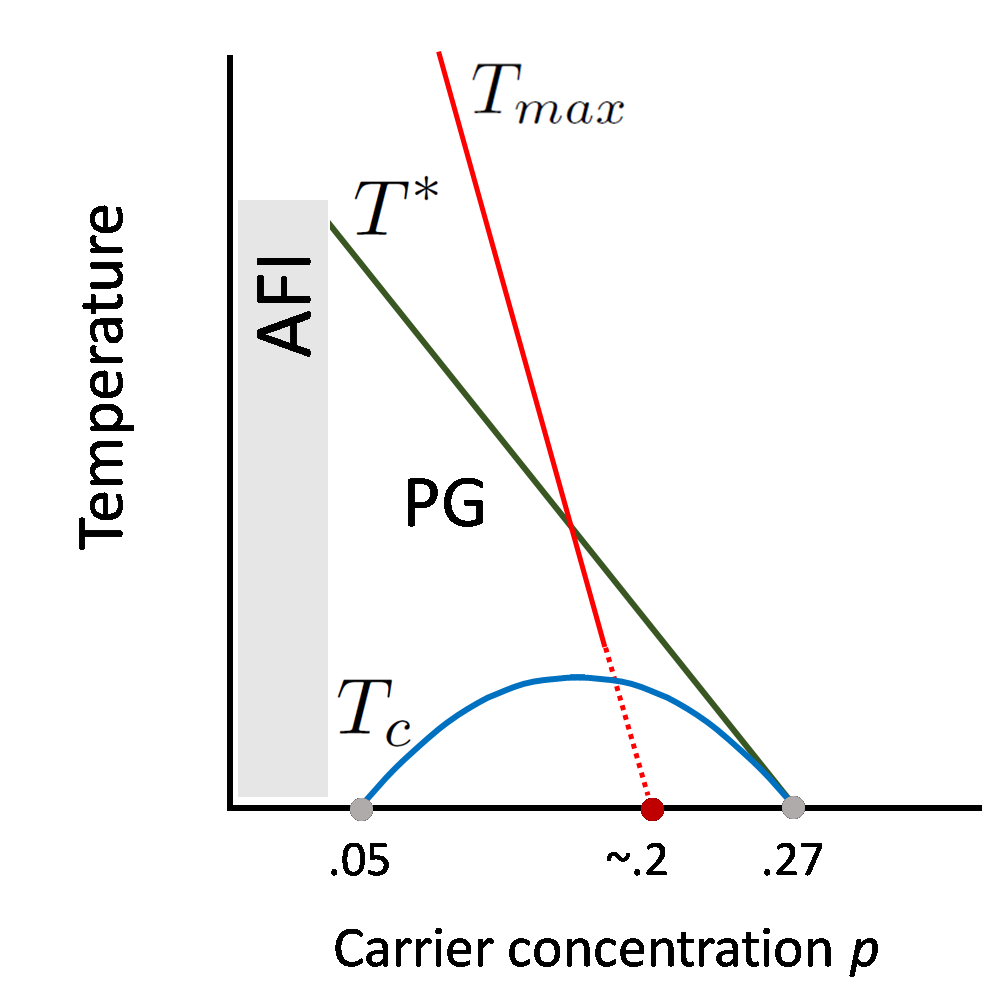}
\caption{(Color online) Essential phase diagram of cuprates (see \cite{Article_Chi} for details) with the three characteristic
temperature lines: the critical temperature $T_c(p)$, the pseudogap line $T^*(p)$ and the characteristic temperature of magnetic
correlations, $T_{max}(p)$.} \label{Fig_1}
\end{figure}

Superconductivity emerges in cuprates from doping an antiferromagnetic (AF) Mott insulator, leading to a complex phase
diagram as shown in Fig. \ref{Fig_1}. The latter displays three important lines : the superconducting dome $T_c(p)$, the pseudogap
line $T^*(p)$ and the characteristic temperature of magnetic correlations $T_{max}(p)$.

Despite the very large variety of compounds belonging to the cuprate family, there is a striking universality in the phase diagram. In addition, as already noted by Tahir et al. \cite{JPhysChem_Tahir2010}, the doping values characterizing the
$T_c$-dome ($p_{min}$, the onset doping for superconductivity, $p_{opt}$ the top of the dome or optimal doping and $p_{max}$ the
end of the dome) seem to be universal values, which are practically independent of the material.

All these experimental facts strongly suggest that superconductivity in cuprates is essentially governed by topological constraints imposed on the system of holes by the 2d antiferromagnetic square lattice.

\subsection{Magnetic properties}

Contrary to conventional superconductors which are mostly `good' metals, parent compounds of SC cuprates are antiferromagnetic Mott
insulators, characterized by the temperature of long-range magnetic ordering $T_\textrm{\tiny N\'eel}(p)$. Electron or hole doping
strongly modifies the electronic and magnetic properties. First, the N\'eel temperature decreases rapidly with doping and finally
vanishes for a small value $p\approx$0.015 \cite{PRB_Keimer1992}. Then, above $p_{min}=$0.05, the system becomes metallic and exhibits superconducting properties below the critical temperature $T_c$.

Measurements of the magnetic susceptibility $\chi(T)$ \cite{PRL_Johnston1989,PRB_Torrance1989,PhysicaC_Yoshizaki1990,PhysicaC_Oda1991,PRB_Nakano1994} have shown that the characteristic temperature of magnetic correlations, $T_{max}$(defined as the peak in the magnetic susceptibility as a function of temperature) decreases with doping. In a previous work \cite{Article_Chi}, we have shown that a similar behavior is observed in the four different compounds (La$_{2-x}$Sr$_x$CuO$_4$, Bi$_2$Sr$_2$Ca$_{1-x}$Y$_x$Cu$_2$O$_8$, Bi$_2$Sr$_2$CaCu$_2$O$_{8+y}$, YBa$_2$Cu$_3$O$_{6+y}$): $T_{max}(p)$ decreases linearly with $p$ over a wide range and then saturates in the overdoped regime. In addition, the extrapolation of the linear behavior to zero temperature gives a critical doping $p_c\sim$0.2.

\begin{figure}
\hspace*{-5mm}\includegraphics[width=9.5 cm]{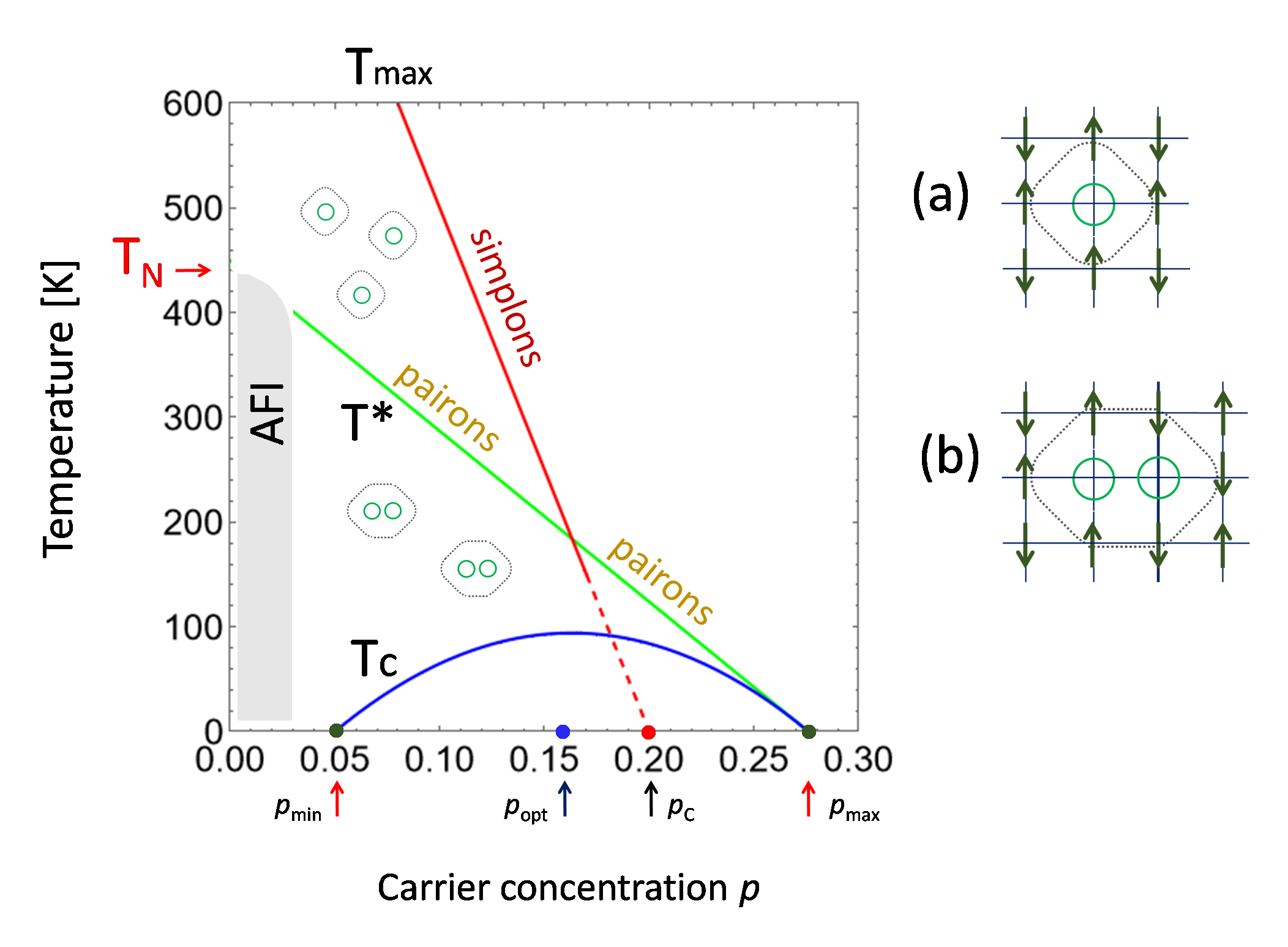}
\caption{(Color online) Phase diagram in the pairon-simplon viewpoint. In the underdoped regime, a simple picture emerges: simplons (right panel, a) exist below the line $T_{max} (p)$ while incoherent pairons (right panel, b) exist below the pseudogap line $T^*(p)$. Finally, pairons condense below $T_c(p)$ in the coherent SC state. Three key doping points are illustrated: $p_{min}$, the onset doping of superconductivity, $p_{max}$, the end of the $T_c$-dome, and $p_c\sim$0.2 the extrapolated value  of the line $T_{max} (p)$ at $T=0$.} \label{Fig_2}
\end{figure}

\begin{figure}
\hspace*{+1 cm}\includegraphics[width=8.4 cm]{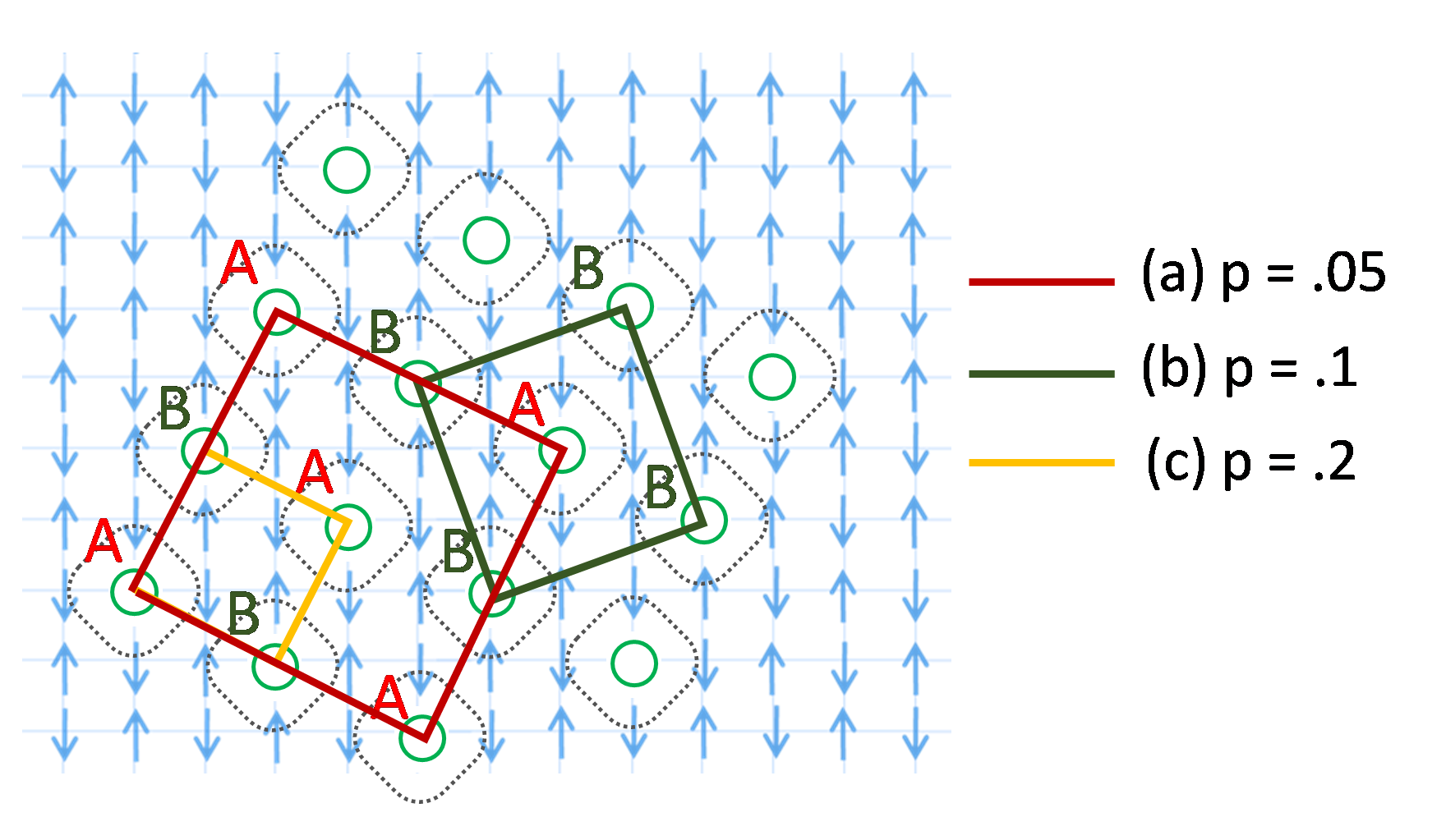}
\caption{(Color online) The compact simplon lattice (yellow square) with superlattice constant $d=\sqrt{5}$, corresponds to the `critical' doping value $p_c=$0.2. The two sublattices of equivalent simplons (A or B, green square) have a lattice constant  $d_0=\sqrt{10}$, giving the doping value $p=$0.1. The doping $p_{min}=$0.05 corresponds to a simplon on the corner of the red (non-centered) square. We use $a_0$ units throughout.} \label{Fig_3}
\end{figure}

The linear behavior of  $T_{max}(p)$ with hole density can be qualitatively understood by considering that each hole `suppresses'
the four nearest neighbors spins. We call this dressed hole a `simplon', which is an effective particle defined as a hole with 4 associated neighboring sites, thus giving 4+1 `spinless' sites (see Fig. \ref{Fig_2}, right
panel). A simple density argument leads to the linear law
\begin{equation}
k_BT_{max}\simeq J(1-5p)
\label{Equa_Tmax}
\end{equation}
where $J$ is the magnetic exchange energy.  The critical doping value $p_c=$0.2, where $T_{max}$ vanishes, corresponds to the
compact simplon lattice (Fig. \ref{Fig_3}), with a superlattice constant $d=\sqrt{5}$. Interestingly, this doping value is very
close to the one which has been identified either as the end of the pseudogap \cite{PhysicaC_Tallon2001} or a quantum critical point (QCP) \cite{AnRevCondMat_Proust2019}.

Two types of simplons, A and B, can be distinguished depending on the spin state of the four electrons surrounding a hole.
The unit cell of the square sublattice of equivalent simplons A (or B) has a side $d=\sqrt{10}$, corresponding to a density $p=$0.1 (see Fig. \ref{Fig_3}). The percolation point of equivalent simplons on such a sublattice is the previous density divided by two, i.e. $p=$0.05. This is very close to the experimental value of $p_{min}$, the onset doping for superconductivity. Metallicity could arises preferentially due to the coupling of equivalent simplons (A or B), which can interact provided their distance is smaller than  $d_0\approx\sqrt{20}$.

\subsection{Pairing in cuprates}

In conventional superconductors, a bound state of electrons can form in the electron sea as a result of phonon exchange. Cooper
pairs \cite{PR_Cooper1956} are delocalized objects which are well described in $k$-space as pairs of opposite wave vectors and spins, $\left|\vec{k}\uparrow -\vec{k}\downarrow \right\rangle$.

We proposed in Ref. \cite{EPL_Sacks2017} that the pairing  mechanism in cuprate belongs to a completely different class. The magnetic ordering of the N\'eel state is destroyed by hole doping but survives on the local scale. As shown by Birgeneau et al. by neutron measurements in Ref. \cite{PRB_Birgeneau1988}, the antiferromagnetic
coherence length varies roughly as the average distance between hole $\xi_{AF}\sim a/\sqrt{p}$. In our model
\cite{EPL_Sacks2017,EPL_Noat2019}, below the characteristic temperature $T^*$, two adjacent holes tend to form a bound state due to this local AF environment, with a binding energy on the order of $J$, as confirmed by early numerical calculations
with the Hubbard or t-J hamiltonian \cite{PRB_Kaxiras1988,PRB_Bonca1989,PRB_Riera1989,PRB_Hasegawa1989,PRB_Poilblanc1994}.
Thus, a new kind of pair exists in cuprates, pairs of holes or `pairons' in their local AF environment, in real space on a typical
length $\xi_{AF}$. The existence of pairons is supported by the detailed angular dependence of the gap function \cite{EPL_Noat2019}
measured by Angular Resolved Photoemission Spectroscopy (ARPES) \cite{PRL_Terashima2007,Natcom_Anzai2013}.

Electrons within the AF coherence length contribute to the pairing energy of the two holes. As the doping increases, the AF coherence length decreases as $\sim 1/\sqrt{p}.$ A pairon can exist provided two adjacent holes are surrounded by at least a ring of six electrons. The pairons thus occupy a minimum of 8 sites (2 holes plus six electrons) on the square lattice in the local AF
environnement.

\begin{figure}
\hspace*{-1.0 cm}\includegraphics[width=10.0 cm]{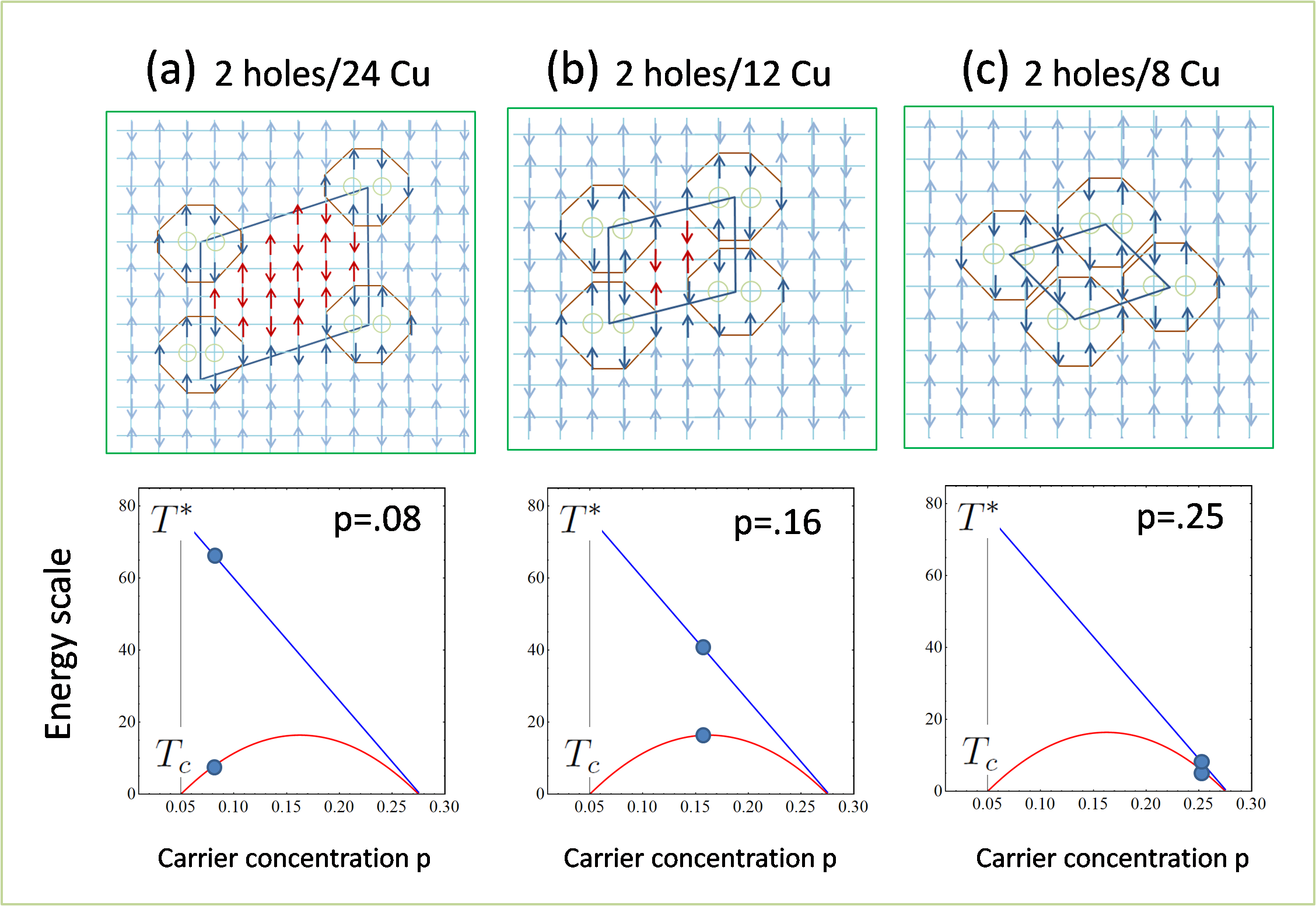}
\caption{(Color online) Upper panel: Schematic of the pairon lattice for three different doping values corresponding to a) underdoped regime, b) optimal doping, c) maximum doping (compact pairon
lattice).} \label{Fig_4}
\end{figure}

This topological constraint imposes a maximum doping value corresponding to the compact pairon lattice (see Fig. \ref{Fig_4}),
which is obtained for p=1/4. This value is remarkably close to the observed $p_{max}=$0.27 at the dome limit. Actually, experimental facts indicate that pairons can still exist for a slightly larger doping value 0.25$+\delta$, up to $p_{max}=$0.27. The reason for this small excess is unknown and requires further thought, but the experimental value corresponds to an additional 1/6 hole per pairon ring ($\delta=x/8=1/48\approx$0.02). Nevertheless, the compact pairon lattice provides a satisfactory topological interpretation.

As previously mentioned, the distance between pairons is of the order of the spin correlation length $\xi_{AF}$. The hypothesis that only sites within the correlation length contribute to the pairon binding energy implies that the binding energy, or equivalently the temperature of pairon formation $T^*$, varies linearly as a function of doping \cite{EPL_Sacks2017}:
\begin{equation}
T^* \propto (1-4p) .
\label{Equa_Tstar}
\end{equation}
This linear law is accurately confirmed by ARPES \cite{RepProgPhys_Hufner2008,Nat_Hashimoto2014} and tunneling
measurements \cite{JPhysSocJap_Nakano1998,Revmod_Fisher2007}.

Both the $T_{max}(p)$ and $T^*(p)$ lines are thus given by similar topological arguments based on two fundamental quantum objects,
simplons and pairons.

\subsection{Superconducting condensation}

In conventional superconductors, the gap $\Delta(T)$ is the order parameter \cite{PR_BCS1957}. At finite temperature, the SC state is gradually destroyed by means of quasiparticle fermionic excitations: the pair-breaking mechanism. Consequently, the gap  $\Delta(T)$ decreases with increasing temperature and finally vanishes at the critical temperature $T_c$. These considerations are captured by the BCS relation \cite{PR_BCS1957}: $1.7k_BT_c=\Delta(0)$.

The situation must be different in cuprates since, as well demonstrated unequivocally by tunneling spectroscopy
\cite{PRL_renner1998_T,JphysSocJap_Sekine2016} and ARPES \cite{Nat_Ding1996,Nat_Hashimoto2014} measurements, the gap clearly
does not vanish at $T_c$ and a pseudogap remains up to the higher temperature $T^*$. The conventional BCS relation is therefore no
longer valid. What is then the order parameter in cuprates?

In several articles \cite{SciTech_Sacks2015,EPL_Sacks2017}, we have proposed that, unlike conventional SC, condensation in cuprates
arises because of pairon-pairon interactions. Moreover, the fundamental excitations of the condensate are pairon excitations
governed by Bose statistics \cite{Jphys_Sacks2018,SolStatCom_Noat2021}. A simple picture emerges (see Fig. \ref{Fig_2}): $T_{max}$ corresponds to the characteristic temperature of magnetic correlations below which simplons are
formed. Below $T^*$ pairons are formed, corresponding to the pseudogap state where they remain incoherent. Finally, below $T_c$,
pairons condense in the coherent SC state.

Therefore, in our model, the total energy $E_{SC}$ of the superconducting state (per pair) is unconventional and reads
\cite{SolStatCom_Noat2021}:
\begin{equation}
E_{SC}=-\Delta_p-\beta_c
\label{Equa_Esc}
\end{equation}
where  $\Delta_p$ is the zero-temperature binding energy and $\beta_c$ is the condensation energy responsible for long range
order. The latter can be precisely extracted from the experimental quasiparticle spectra \cite{PRB_Sacks2006,SciTech_Sacks2015}. Contrary to BCS, since $\Delta_p$ is constant across $T_c$, it is $\beta_c$ that determines the critical temperature and not the gap \cite{SciTech_Sacks2015}, with
the result:
\begin{equation}
\beta_c\simeq 2.2k_BT_c \label{Equa_eta_Tc}
\end{equation}
We now show that this condensation energy $\beta_c$, i.e. the energy difference between the non SC pseudogap state $E_{PG}=-\Delta_p$ and the SC state $E_{SC}$, results from {\it quantifying the amount of pairon disorder}. In this view, the characteristic disorder in the PG state is uniquely determined by the topological constraints of the 2d square lattice.

\begin{figure}
\hspace*{-1 cm}\includegraphics[width=10.0 cm]{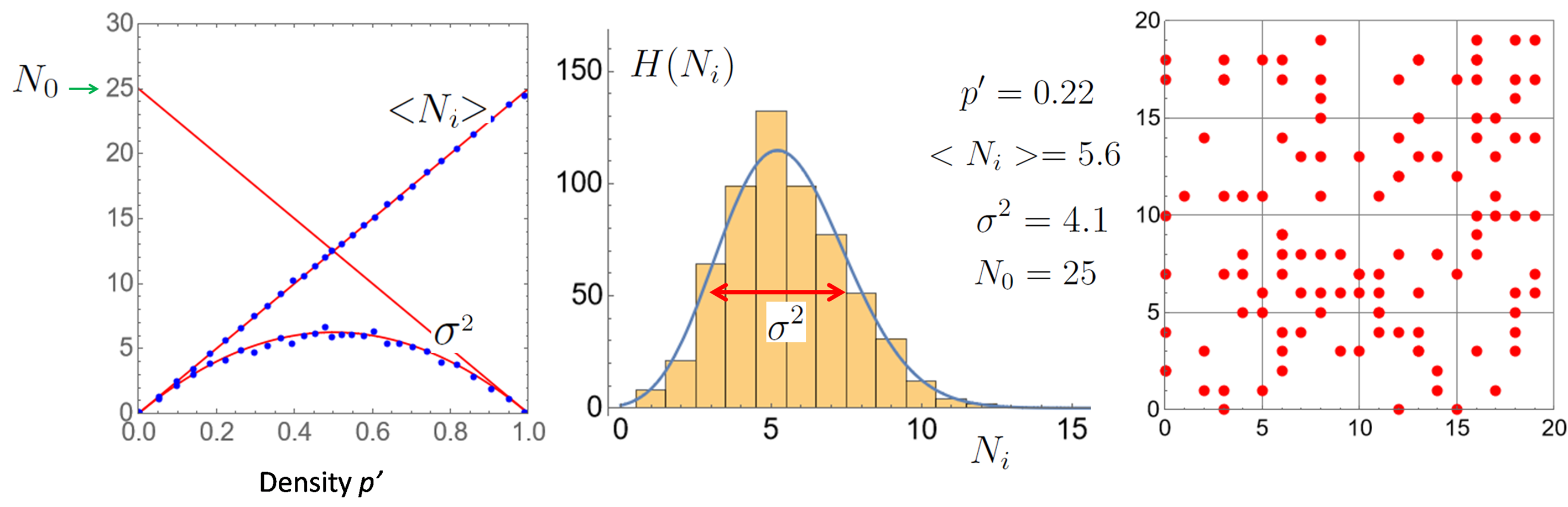}
\caption{(Color online)  Statistics of the number of pairons $N_i$ in a square of  size $d_0=5$ containing $N_0=$25 possible sites.
Left panel : Mean value $\left\langle N_i\right\rangle$ and variance $\sigma^2=\left\langle N_i^2\right\rangle-\left\langle
N_i\right\rangle^2$. plotted as a function of nominal pairon density $p^\prime$. Middle panel : Histogram $H(N_i)$ of the number of pairons $N_i$ for $p^\prime=$0.22, chosen as an example. Right panel: Pairons randomly distributed on the pairon lattice for
$p^\prime=$0.22.} \label{Fig_5}
\end{figure}
To proceed, akin to a lattice gas approach, let us consider pairons on an equivalent square lattice
of density $p^\prime$: $p^\prime=$1 corresponds to the compact pairon lattice ($p=$0.27 on the Cu0 square lattice) and $p^\prime=$0
to the onset doping at which the average distance between pairons is small enough ($d<d_0$) so that they interact
($p=$.05 on the Cu0 square lattice). Pairons are then randomly distributed on this lattice (Fig. \ref{Fig_5}), under the hypothesis where all sites are equivalent. We calculate the number of pairons, $N_i$, inside a square of side $d_0$ (with $N_0={d_0}^2$ sites) , where $d_0$ is the maximum interacting distance between pairons. From the distribution of $\left\langle N_i\right\rangle$, we calculate the statistical averages, the mean value $\left\langle N_i\right\rangle$
and the variance $\sigma^2=\left\langle N_i^2\right\rangle-\left\langle N_i\right\rangle^2$.

As expected for a problem depending only on the success $p^\prime$ or failure $1-p^\prime$ to find a pairon on a given site, we obtain for $N_i$ a {\it binomial} distribution. As a result, $\left\langle N_i\right\rangle$ follows a straight line as a function of $p^\prime$, with some jitter, while the variance $\sigma^2=N_0 p^\prime(1-p^\prime)$ displays a dome shape. In fact, for a fixed $N_0$ and a random distribution, $\sigma^2$ characterizes the maximum disorder of the localized pairons on the square lattice.

The condensation energy can be understood using the Following gendanken experiment. When the interaction between pairons is switched off, we obtain the incoherent PG state, the `pairon glass', of energy $E_{PG}$, where there is by definition no correlation between pairons. In this
state, no long-range SC order exists and the disorder is described by the above binomial distribution.

When the interaction is turned back on (provided their typical distance $d$ is smaller than $d_0$), all pairons are in the ordered SC ground state, with energy $E_{PG}-\beta_c$. In this zero temperature transformation, the virtual work $W$ needed to disorder the system is $\beta_c$:
\begin{equation}
W=E_{PG}-E_{SC}=\beta_c
\label{Equa_Work}
\end{equation}
Since the disorder is characterised by the variance of the distribution, one should have $W\propto  \sigma^2$. Thus, we obtain
\begin{equation}
\beta_c\propto \sigma^2/N_0=p^\prime(1-p^\prime).
\label{Equa_Beta}
\end{equation}
In this view, the condensation is a new type of disorder to order transition, independent of the other degrees of freedom. In spite of the complex excitations, Bose excitations and quasiparticle fermionic excitations, and magnetic degrees of freedom, the
underlying topological constraints define the fundamental mechanism.

\subsection{Pseudogap and superconductivity}

We now focus on the interplay between the pseudogap and the superconducting phase in cuprate. The pseudogap line is given by $T^*\propto(1-4p)$, which translates to $\sim(1-p^\prime)$ in the pairon sublattice, and that the condensation energy is $\beta_c\sim p^\prime(1-p^\prime)$. Furthermore, the statistical calculation described above leads the two relations:
\begin{eqnarray}
&&T^*=\lambda \,(1-p^\prime)  \nonumber\\
&& T_c=\lambda \,p^\prime(1-p^\prime)
\label{Equa_Tstar_Tc}
\end{eqnarray}
where $\lambda$ has the dimension of temperature. The true doping value is given by the linear tranformation $p=p_{min}+(p_{max}-p_{min})\times p^\prime$.
\begin{figure}
\includegraphics[width=8.4 cm]{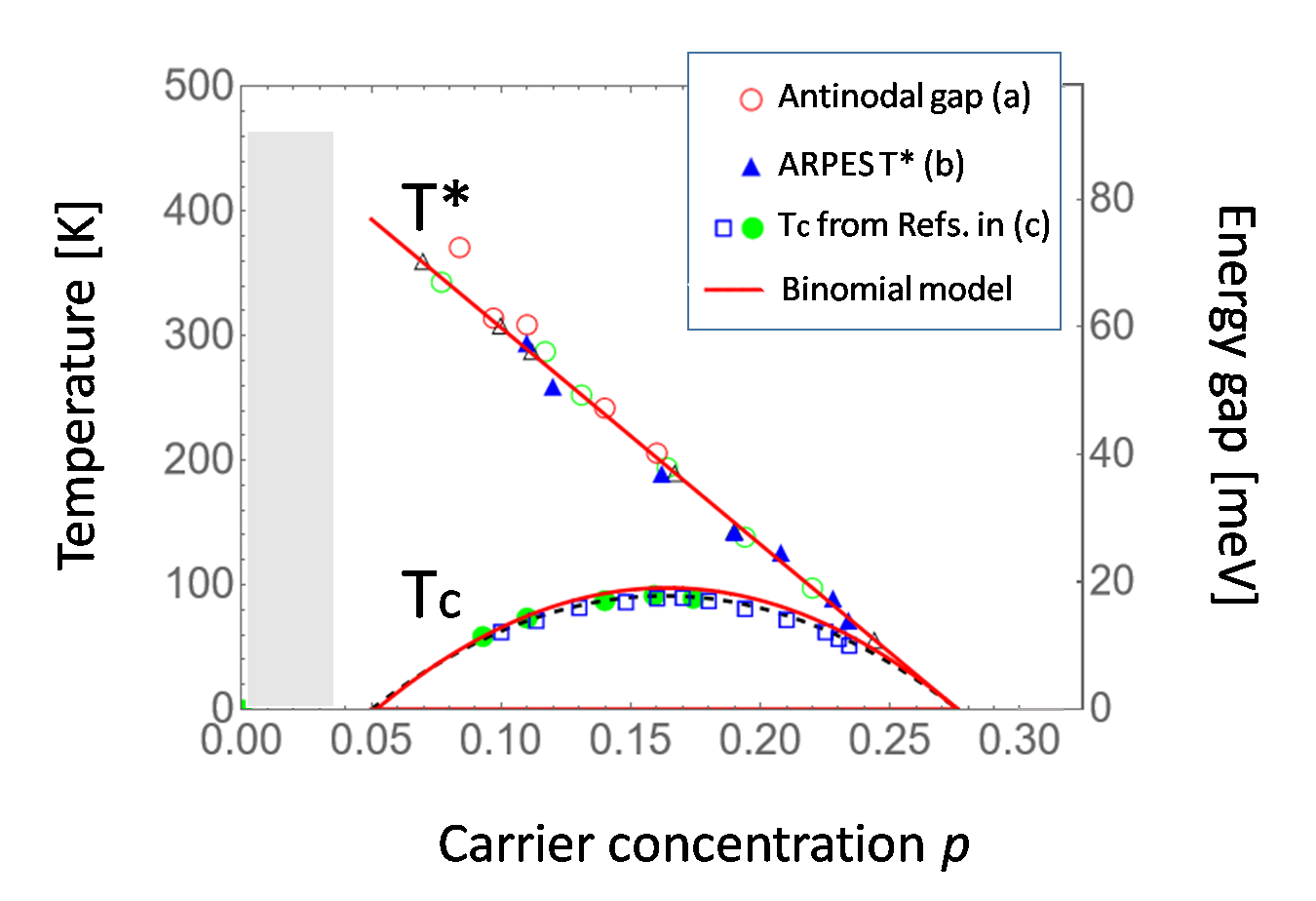}
\caption{(Color online) (a) Antinodal gap (in units of
$\Delta/2.2k_B$) (b) pseudogap temperature $T^*$, both measured by ARPES \cite{Nat_Hashimoto2014, PNAS_Vishik2012} and (c) experimental superconducting temperature $T_c$ (from Ref.
\cite{PhysicaC_Tallon2001}) , compared to the binomial law (red lines) with only one adjustable parameter $\lambda$. From the fit of the data we find the value $\lambda=$400K.} \label{Fig_6}
\end{figure}

To proceed, we now compare in Fig. \ref{Fig_6} the calculated $T^*$ and $T_c$ to the experimental values, measured by ARPES
\cite{PNAS_Vishik2012} and resistivity \cite{PhysicaC_Tallon2001} respectively. With only one adjustable parameter $\lambda$, the
agreement between theory and experiments is remarkable. We find the value $\lambda=$400$K$, which is comparable to $J/2$, or the N\'eel temperature when extrapolated to $p=0$.

The binomial law provides a simple explanation for the maximum $T_c$ obtainable, which occurs at half-filling,
$p'=1/2$. Indeed, at this concentration the fluctuation amplitude, measured by $\sigma^2$, is a maximum. At this optimum concentration, we see the topological constraint that $T^*$ is exactly twice
$T_c$.

These fundamental relations illustrate that the pseudogap and superconductivity are completely linked and arise from the same
physical phenomenon, for the entire doping range. The geometry of the dome and the tangent line are not arbitrary but determined by
the statistics of pairons randomly distributed on a square lattice. A direct consequence is that both $T^*$ and $T_c$ are proportional to the same energy scale, which we identify as the magnetic exchange
energy $J$.

\subsection{Conclusion}

In this article, we propose that key aspects of the phase diagram of cuprates are governed by the topological properties of the doped antiferromagnetic insulator on a square lattice. Bound pairs of
holes are formed due to the local antiferromagnetic environnement and condense in the superconducting state in a {\it disorder to
order} transition. The condensation is driven by a new mechanism and is directly related to the amount of disorder in the non SC
pseudogap state. Thus the simple binomial pairon distribution explains the $T_c$-dome as well as the pseudogap line. We show that
the cuprate SC state, a spatially correlated quantum state of pairons, is intimately connected to the disordered pseudogap state
-- in our view they appear as indissociable phenomena.

\subsection{Acknowledgments}

The authors gratefully acknowledge many fruitful discussions with H.
Eisaki and S. Ishida (AIST, Tsukuba, Japan) and M. Nohara
(University of Hiroshima, Japan).


\vskip 2 mm


\begin{thebibliography}{99}


\bibitem{JPhysChem_Tahir2010} Jamil Tahir-Kheli and William A. Goddard III, Universal Properties of Cuprate Superconductors: Tc Phase Diagram, Room-Temperature Thermopower, Neutron Spin Resonance, and STM Incommensurability Explained in Terms of Chiral Plaquette Pairing, J. Phys. Chem. Lett. {\bf 1}, 1290--1295 (2010).

\bibitem{Article_Chi}  Y. Noat, A. Mauger, M. Nohara, H. Eisaki, W. Sacks, Cuprates phase diagram deduced from magnetic susceptibility: what is the 'true' pseudogap line?,
article submitted to Solid State communications (2022).

\bibitem{PRB_Keimer1992} B. Keimer, N. Belk, R. J. Birgeneau, A. Cassanho, C. Y. Chen, M. Greven, M. A. Kastner, A. Aharony, Y. Endoh, R. W. Erwin, and G. Shirane, Magnetic excitations in pure, lightly doped, and weakly metallic La$_2$CuO$_4$, Phys. Rev. B {\bf 46}, 14034 (1992).

\bibitem{PRL_Johnston1989} David C. Johnston, Magnetic Susceptibility Scaling in La$_{2-x}$Sr$_x$CuO$_{4-y}$, Phys. Rev. Lett. {\bf 62}, 957 (1989).

\bibitem{PRB_Torrance1989} J. B. Torrance, A. Bezinge, A. I. Nazzal, T. C. Huang, S. S. P. Parkin, D. T. Keane, S. J. LaPlaca, P. M. Horn, and G. A. Held, Properties that change as superconductivity disappears at high-doping concentrations in La$_{2-x}$Sr$_x$CuO$_4$, Phys. Rev. B {\bf 40}, 8872 (1989).

\bibitem{PhysicaC_Yoshizaki1990} R. Yoshizaki, N. Ishikawa, H. Sawada, E. Kita, A. Tasaki, Magnetic susceptibility of normal state and superconductivity of La$_{2-x}$Sr$_x$CuO$_4$, Physica C {\bf 166}, 417 (1990).

\bibitem{PhysicaC_Oda1991} M.Oda, T.Nakano, Y.Kamada, M.Ido, Electronic states of doped holes and magnetic properties in La$_{2?x}$M$_x$CuO$_4$ (M = Sr, Ba), Physica C {\bf 183}, 234 (1991).

\bibitem{PRB_Nakano1994} T. Nakano, M. Oda, C. Manabe, N. Momono, Y. Miura, and M. Ido, Magnetic properties and electronic conduction of superconducting  La$_{2-x}$Sr$_x$CuO$_4$, Phys. Rev. B {\bf 49}, 16000 (1994).

\bibitem{PhysicaC_Tallon2001} J. L. Tallon and J.W. Loram, The doping dependence of $T^*$ - what is the real high-$T_c$ phase diagram?, Physica C {\bf 349}, 53 (2001).

\bibitem{AnRevCondMat_Proust2019} C. Proust and L. Taillefer, Annual Review of Condensed Matter Physics {\bf 10}, 409 (2019).

\bibitem{PR_Cooper1956} Leon N. Cooper, Bound electron pairs in a degenerate Fermi gas, Physical Review {\bf 104}, 1189-1190 (1956).

\bibitem{EPL_Sacks2017} W. Sacks, A. Mauger and Y. Noat, Cooper pairs without glue in high-$T_c$ superconductors: A universal phase diagram, Euro. Phys. Lett {\bf 119}, 17001 (2017).

\bibitem{PRB_Birgeneau1988} R. J. Birgeneau, D. R. Gabbe, H. P. Jenssen, M. A. Kastner, P. J. Picone, T. R. Thurston, G. Shirane, Y. Endoh, M. Sato, K. Yamada, Y. Hidaka, M. Oda, Y. Enomoto, M. Suzuki, and T. Murakami, Antiferromagnetic spin correlations in insulating, metallic, and superconducting La$_{2-x}$Sr$_x$CuO$_4$, Phys. Rev. B {\bf 38}, 6614 (1988).

\bibitem{EPL_Noat2019} Y. Noat, A. Mauger and W. Sacks, Single origin of the nodal and antinodal gaps in cuprates,
Euro. Phys. Lett {\bf 126}, 67001 (2019).

\bibitem{PRB_Kaxiras1988} Efthimios Kaxiras and Efstratios Manousakis, Hole dynamics in the two-dimensional strong-coupling Hubbard Hamiltonian, Phys. Rev. B {\bf 38}, 866(R) (1988).

\bibitem{PRB_Bonca1989} J. Bon\v{c}a, P. Prelov\v{s}ek, and I. Sega, Exact-diagonalization study of the effective model for holes in the planar antiferromagnet, Phys. Rev. B {\bf 39}, 7074 (1989).

\bibitem{PRB_Riera1989} J. A. Riera and A. P. Young, Binding of holes in one-band models of oxide superconductors, Phys. Rev. B {\bf 39}, 9697(R) (1989).

\bibitem{PRB_Hasegawa1989} Y. Hasegawa and D. Poilblanc, Hole dynamics in the t-J model: An exact diagonalization study, Phys. Rev. B {\bf 40}, 9035 (1989).

\bibitem{PRB_Poilblanc1994} Didier Poilblanc, Jos\'e Riera and Elbio Dagotto, d-wave bound state of holes in an antiferromagnet, Phys. Rev. B {\bf 49}, 12318 (1994).


\bibitem{PRL_Terashima2007} K. Terashima, H. Matsui, T. Sato, T. Takahashi, M. Kofu, and K. Hirota, Anomalous Momentum Dependence of the Superconducting Coherence Peak and Its Relation to the Pseudogap of La$_{1.85}$Sr$_{0.15}$CuO$_{4}$,
Phys. Rev. Lett. {\bf 99}, 017003 (2007).

\bibitem{Natcom_Anzai2013} H. Anzai, A. Ino, M. Arita, H. Namatame, M. Taniguchi, M. Ishikado, K. Fujita, S. Ishida and S. Uchida, Relation between the nodal and antinodal gap and critical temperature in superconducting Bi2212, Nature Communications {\bf 4}, 1815 (2013).

\bibitem{RepProgPhys_Hufner2008} S. H\"ufner, M. A. Hossain, A Damascelli, and G. A. Sawatzky,Two gaps make a high-temperature superconductor?, Rep. Prog. Phys., {\bf 71}, 062501 (2008).

\bibitem{Nat_Hashimoto2014} Makoto Hashimoto, Inna M. Vishik, Rui-Hua He, Thomas P. Devereaux and Zhi-Xun Shen, Energy gaps in high-transition-temperature cuprate superconductors, Nature Physics {\bf 10}, 483 (2014).

\bibitem{JPhysSocJap_Nakano1998} Tohru Nakano, Naoki Momono, Migaku Oda, and Masayuki Ido, Correlation between the Doping Dependences of Superconducting Gap Magnitude $2\Delta_0$ and Pseudogap Temperature $T^*$ in High--$T_c$ Cuprates, J. Phys. Soc. Jpn. {\bf  67},  2622 (1998).

\bibitem{Revmod_Fisher2007} \O. Fischer, M. Kugler, I. Maggio-Aprile,
C. Berthod and C. Renner, Scanning tunneling spectroscopy of the
cuprates, Rev. Mod. Phys. {\bf 79}, 353 (2007).

\bibitem{PR_BCS1957} J. Bardeen, L. Cooper, J. Schrieffer, Theory of Superconductivity, Phys. Rev. {\bf 108} 1175 (1957).

\bibitem{PRL_renner1998_T} Ch. Renner,B. Revaz, J.-Y. Genoud, K. Kadowaki,and {{\O}}. Fischer, Pseudogap precursor of the superconducting gap in under- and overdoped Bi$_2$Sr$_2$CaCu$_2$O$_{8+\delta}$, Phys. Rev. Lett., {\bf 80} 149 (1998).

\bibitem{JphysSocJap_Sekine2016} R. Sekine, S. J. Denholme, A. Tsukada, S. Kawashima, M. Minematsu,T. Inose, S. Mikusu, K. Tokiwa, T. Watanabe, and N. Miyakawa, Characteristic features of the mode energy estimated from tunneling conductance on TlBa$_2$Ca$_2$Cu$_3$O$_{8.5+\delta}$, J. Phys. Soc. Jpn. {\bf 85}, 024702 (2016).

\bibitem{Nat_Ding1996} H. Ding, T. Yokoya, J. C. Campuzano, T. Takahashi, M. Randeria, M. R. Norman, T. Mochiku, K. Kadowaki and J. Giapintzakis, Spectroscopic evidence for a pseudogap in the normal state of underdoped high--$T_c$ superconductors,
Nature {\bf 382}, 51 (1996).

\bibitem{SciTech_Sacks2015}  W. Sacks, A. Mauger, Y. Noat, Pair\,--\,pair interactions as a mechanism for
high-T$_c$ superconductivity, Superconduct. Sci. Technol., {\bf 28}
105014 (2015).

\bibitem{Jphys_Sacks2018}W. Sacks, A. Mauger and Y. Noat, Origin of the Fermi arcs in cuprates: a dual role of quasiparticle and pair excitations, Journal of Physics: Condensed Matter, {\bf 30},  475703 (2018).

\bibitem{SolStatCom_Noat2021} Y. Noat, A. Mauger, M. Nohara, H. Eisaki, W. Sacks
, How `pairons' are revealed in the electronic specific heat of cuprates, Solid State Communications {\bf 323}, 114109 (2021).

\bibitem{PRB_Sacks2006} W. Sacks, T. Cren, D. Roditchev, and B. Dou\c{c}ot,Quasiparticle spectrum of the cuprate Bi$_2$Sr$_2$CaCu$_2$O$_{8+\delta}$: Possible connection to the phase diagram, Phys. Rev. B {\bf 74}, 174517 (2006).

\bibitem{PNAS_Vishik2012} I. M. Vishik, M. Hashimoto, R.-H. He, W.-S. Lee, F. Schmitt, D. Lu, R. G. Moore, C. Zhang, W. Meevasana, T. Sasagawa, S. Uchida, Kazuhiro Fujita, S. Ishida, M. Ishikado, Y. Yoshida, H. Eisaki, Z. Hussain, T. P. Devereaux, and Z.-X. Shen, Phase competition in trisected superconducting dome, PNAS {\bf 109}, 18332 (2012).

\end{thebibliography}
\end{document}